\documentclass[10pt, conference, letterpaper]{IEEEtran}

\usepackage{hyperref}
\usepackage{array}
\usepackage{graphicx}
\usepackage[table]{xcolor}
\usepackage{algpseudocode}
\usepackage{hyperref}

\newcolumntype{C}[1]{>{\centering\let\newline\\\arraybackslash\hspace{0pt}}m{#1}}
\algnewcommand\algorithmicinput{\textbf{Input:}}
\algnewcommand\INPUT{\item[\algorithmicinput]}



\title{Multi-Hop Communication for nTorrent in a Wireless Ad Hoc Environment}
\author{
\IEEEauthorblockN{Kimberly Chou}
\IEEEauthorblockA{University of California, Los Angeles \\  klchou@cs.ucla.edu}
}

\usepackage{etoolbox}
\usepackage{xstring}
\DeclareListParser{\doslashlist}{/}
\newcounter{ndnNameComponentCounter}%
\newcommand{\ndnName}[1]{{%
  \setcounter{ndnNameComponentCounter}{0}%
  \renewcommand{\do}[1]{{%
    \ifnumgreater{\value{ndnNameComponentCounter}}{0}{\allowbreak/}{}%
    \ifnumodd{\value{ndnNameComponentCounter}}{}{}%
    \detokenize{##1}}%
    \stepcounter{ndnNameComponentCounter}}%
``{\fontfamily{cmtt}\small\selectfont\IfBeginWith{#1}{/}{/}{}\doslashlist{#1}}''%
}}

\usepackage{eso-pic,xcolor}
\makeatletter
\AddToShipoutPicture*{%
\setlength{\@tempdimb}{20pt}%
\setlength{\@tempdimc}{\paperheight}%
\setlength{\unitlength}{1pt}%
}
\makeatother

\usepackage{color,soul}
\usepackage{xspace}

\begin{document}
\maketitle

\begin{abstract}

nTorrent is a BitTorrent-like application that is based on NDN (Named Data Networking). Ad hoc environments introduce additional challenges to the dissemination of files among peers. Some issues that we encounter are that not all peers in the neighborhood or environment run the nTorrent application or desire the same torrent file. These issues cause nTorrent interests to be unable to be processed or prevent peers from downloading their desired torrent files. In order to solve this issue, I implemented pure forwarding nodes that represent peers that do not run the nTorrent application and also extended the original nTorrent application to be able to forward interests for torrent files other than their own desired torrent file. For this project, the solution is able to facilitate multi-hop communication through all nodes present in the environment whether or not they run nTorrent.

\end{abstract}

\section{Introduction}

Named Data Networking (NDN)~\cite{zhang2010named} is a proposed Internet architecture that makes data the focus of the communication paradigm. nTorrent~\cite{mastorakis2017ntorrent} is peer-to-peer file sharing application, like BitTorrent but uses NDN instead of the IP protocol. nTorrent has been ported to ndnSIM~\cite{mastorakis2017evolution, mastorakis2016ndnsim} with examples where nodes are placed in a wireless ad hoc environment and transfer files to each other. To mimic more realistic environments, some nodes in the environment do not understand the nTorrent application and other nTorrent nodes want to download different torrent files from each other. The main goal of this paper is to support multi-hop communication between all the nodes in the environment.

The code is available at \url{https://github.com/kchou1/scenario-ntorrent}.

\section {Background}

\begin{figure}[h]
  \centering
  \includegraphics[width=\columnwidth]{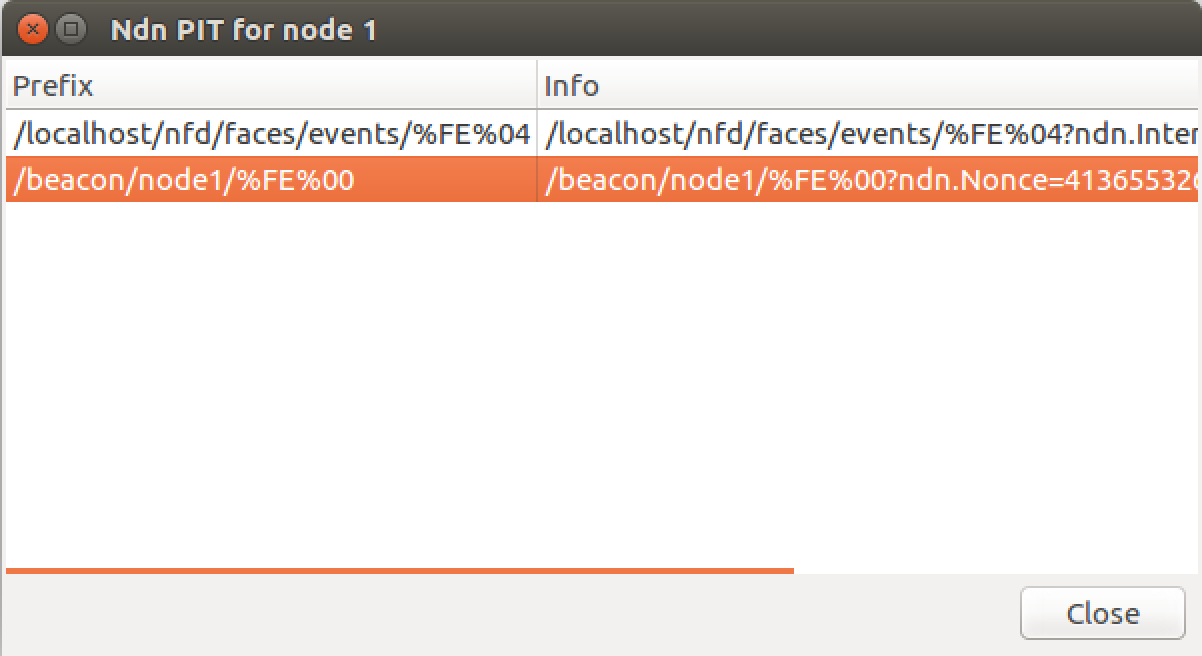}
  \caption{\small Beacons sent}
  \label{Figure:lone-peer}
\end{figure}

\begin{figure}[h]
  \centering
  \includegraphics[width=\columnwidth]{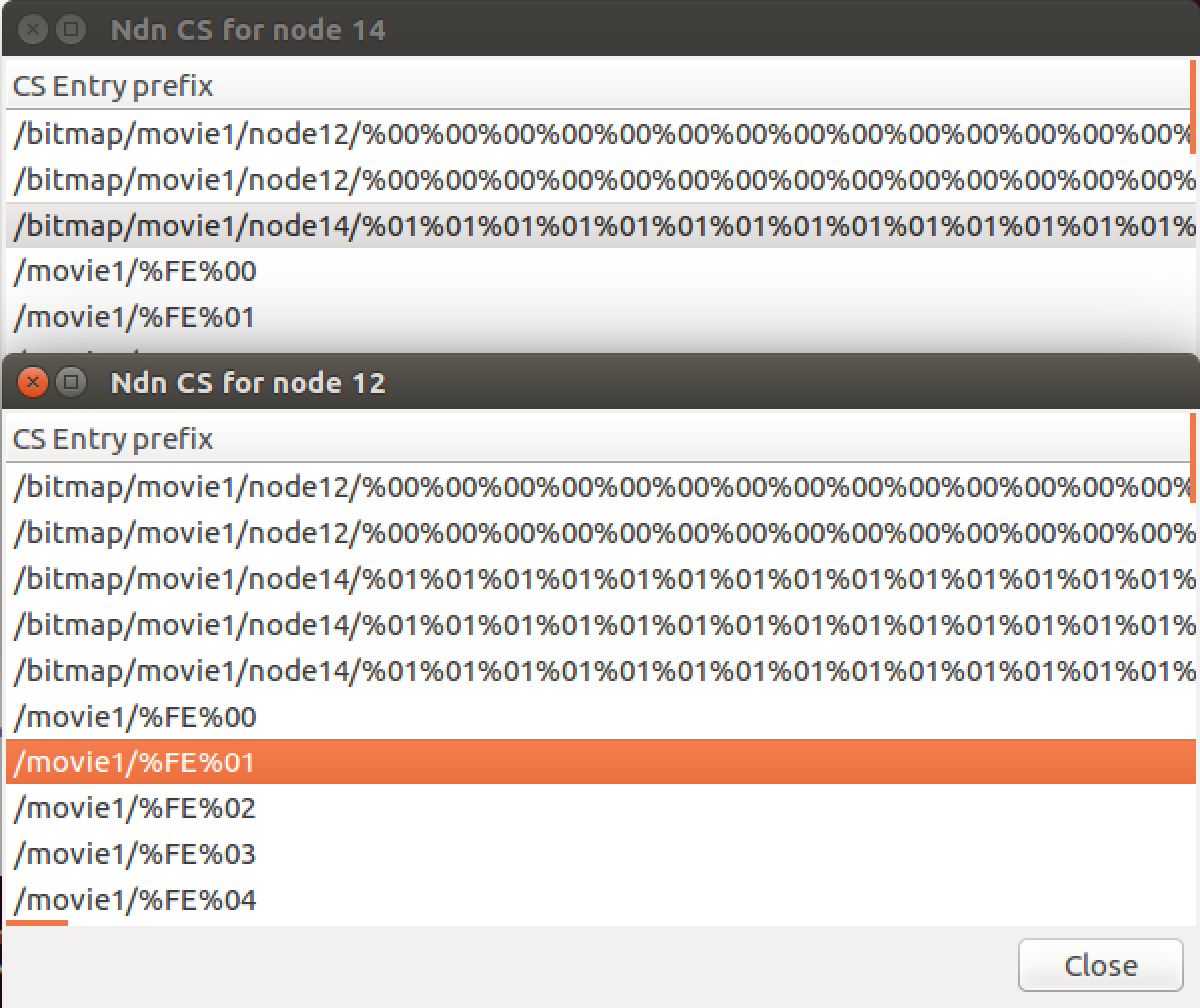}
  \caption{\small Bitmaps exchanged and torrent file packets received by peers}
  \label{Figure:bitmap-exchange}
\end{figure}

In a wireless ad hoc environment, peers running nTorrent use beacons (Figure~\ref{Figure:lone-peer}) and bitmaps (Figure~\ref{Figure:bitmap-exchange}) to communicate with one another about what pieces of the torrent file they have. Beacons would be used to let nearby peers know that the sender of the beacon is close and ready to exchange bitmaps. Each peer’s bitmap contains information about what pieces of the torrent file that peer has. During the exchange of bitmaps, peers figure out which pieces they need and send out Interests to request them. In previous implementations and simulation scenarios, there was always one torrent file that the peers wanted to download. In order to mimic more realistic environments in this project, not all peers want to download the same torrent file and some peers do not run the nTorrent application.

\section {Related Work}

The closest work to this report is nTorrent in wired scenarios~\cite{mastorakis2017ntorrent, mastorakis2017torrent} and a design for distributed dataset synchronization in mobile ad-hoc environments~\cite{li2018data}. This is another project that relied on ndnSIM~\cite{mastorakis2017evolution, mastorakis2016ndnsim, mastorakis2015ndnsim} for evaluation. ndnSIM currently includes integration with the NDN Forwarding Daemon (NFD)~\cite{nfd-dev} and supports the simulation of real-world applications based on the ndn-cxx library and ndnSIM-specific applications based on the application abstraction of NS-3. An example of the former case is RoundSync~\cite{de2017design}, a protocol for distributed data synchronization. Examples of the latter case are KITE~\cite{zhang2018kite} (a framework for producer mobility in NDN), RDR~\cite{mastorakis2018real} (a protocol for the retrieval of real-time data), FIF~\cite{chan2017fuzzy, mastorakis2018experimentation} (a fuzzy Interest forwarding protocol), BELRP~\cite{vusirikala2016hop} (a protocol for best-effort link reliability), and NDN security schemes~\cite{zhang2018security, zhang2018overview}.
\section{Design Overview}

In this section, we describe the high-level design logic for the multi-hop communication and the implementation details.

\subsection{Design}
For this multi-hop communication design, we introduced a peer application that does not understand nTorrent protocol but still understands the NDN network layer semantics. These peers called pure forwarding nodes either forward or do not forward received Interests based on a selected probability. This probability can be configured depending on the simulation or situation. To avoid collisions and flooding the network, peers wait for a certain amount of time before forwarding any Interest. This behavior would still allow the peer to be able to help other peers running nTorrent download their torrent files if they forwarding Interest packets.

nTorrent peers that receive bitmaps and data interests for the same torrent file they want to download will follow the original application. They will receive beacons and send out bitmaps to retrieve the data they want. In order to handle bitmaps for other torrent files, we extend the application to forward these bitmaps or data interests for other torrent files if the peer has overheard other interests for that same torrent file. If the peer has not heard any interests for that torrent file, it will not forward the received interest, but it will remember the torrent file name for a certain amount of time for future interests the peer might hear. So if the peer has heard the torrent file name within a certain amount of time, it can forward the interest because it knows that some other peer is interested in downloading the torrent file that interest is related to. In case peers move away from each other and lose communication, the amount of time set is to avoid flooding the environment with interests and constantly forwarding interests for peers that have moved away.

\begin{figure}[h]
  \centering
  \includegraphics[width=\columnwidth]{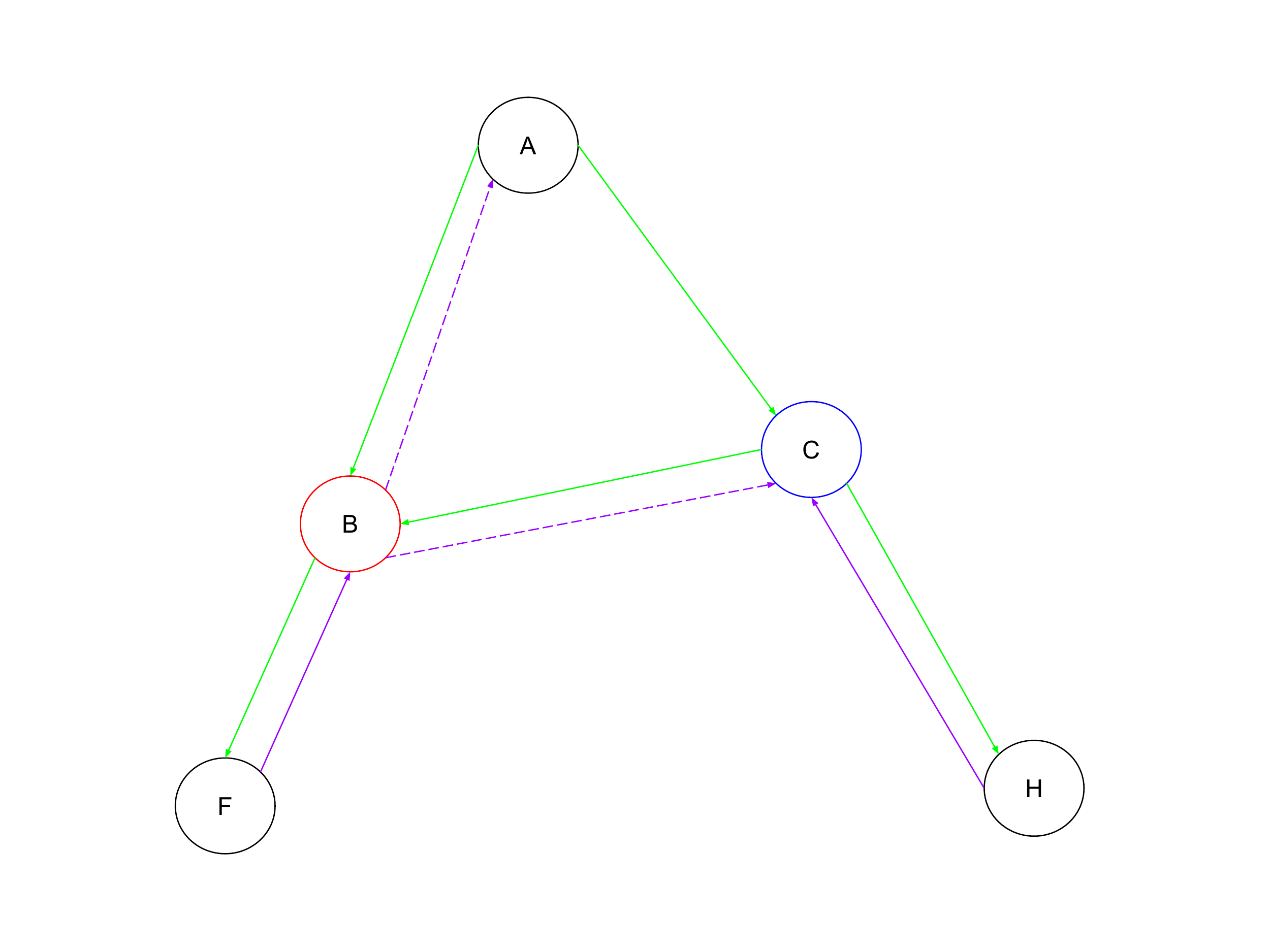}
  \caption{\small Multi-Hop Communication between peers and pure forwarders.}
  \label{Figure:design}
\end{figure}

In Figure~\ref{Figure:design}, node A has the complete set of data packets for one torrent file and node C has the complete set for another. Nodes F and H want to download the torrent file that node A has. Node F must communicate with and through the pure forwarding node, Node B, which will forward interests based on the chosen probability. Node H will communicate with Node C, who will be able to help facilitate the communication to node H due to its knowledge of the nTorrent application protocol.

\section{Implementation}

Using the nTorrent scenario template~\cite{ntorrent-adhoc-repo, ntorrent-code} for ad hoc peers, we created the pure forwarding nodes to forward all Interests received based on a chosen probability that can be configured. At the start we implemented the pure forwarding node as an application, but ran into issues with the forwarding of interests. While running different scenarios, the pure forwarder was not actually forwarding interests to other peers. After some debugging and investigation, we found out that the pure forwarding nodes were not able to forward interests because they would not be sent out on the correct face. So we decided to implement the pure forwarding nodes as a forwarding strategy rather than an application. We still had the pure forwarding nodes forward interests based on a probability, but it would forward the interest to all of its faces in the simulation. This way it would be able to get the interest to other nodes instead of getting stuck at itself.

To test all parts of the implementation, we have one pure forwarding node, one peer that wants to download movie1, one peer that wants to download movie2, one peer that has movie1, and one peer that has movie2. All of the peers except for the pure forwarding node will follow the nTorrent application protocol and communicate with each other. If the peer that wants movie2 receives an interest related to movie1 for the first time, it will keep an entry for the torrent file name and set a expiration time for it. The next time it receives an interest related to movie1, it will forward it because it knows that some other node also desires that torrent file. With this, communication between the peers is able to flow through the pure forwarding node and peers that want different torrent files allowing each peer to finish downloading its desired file.

\begin{figure}[h]
  \centering
  \includegraphics[width=\columnwidth]{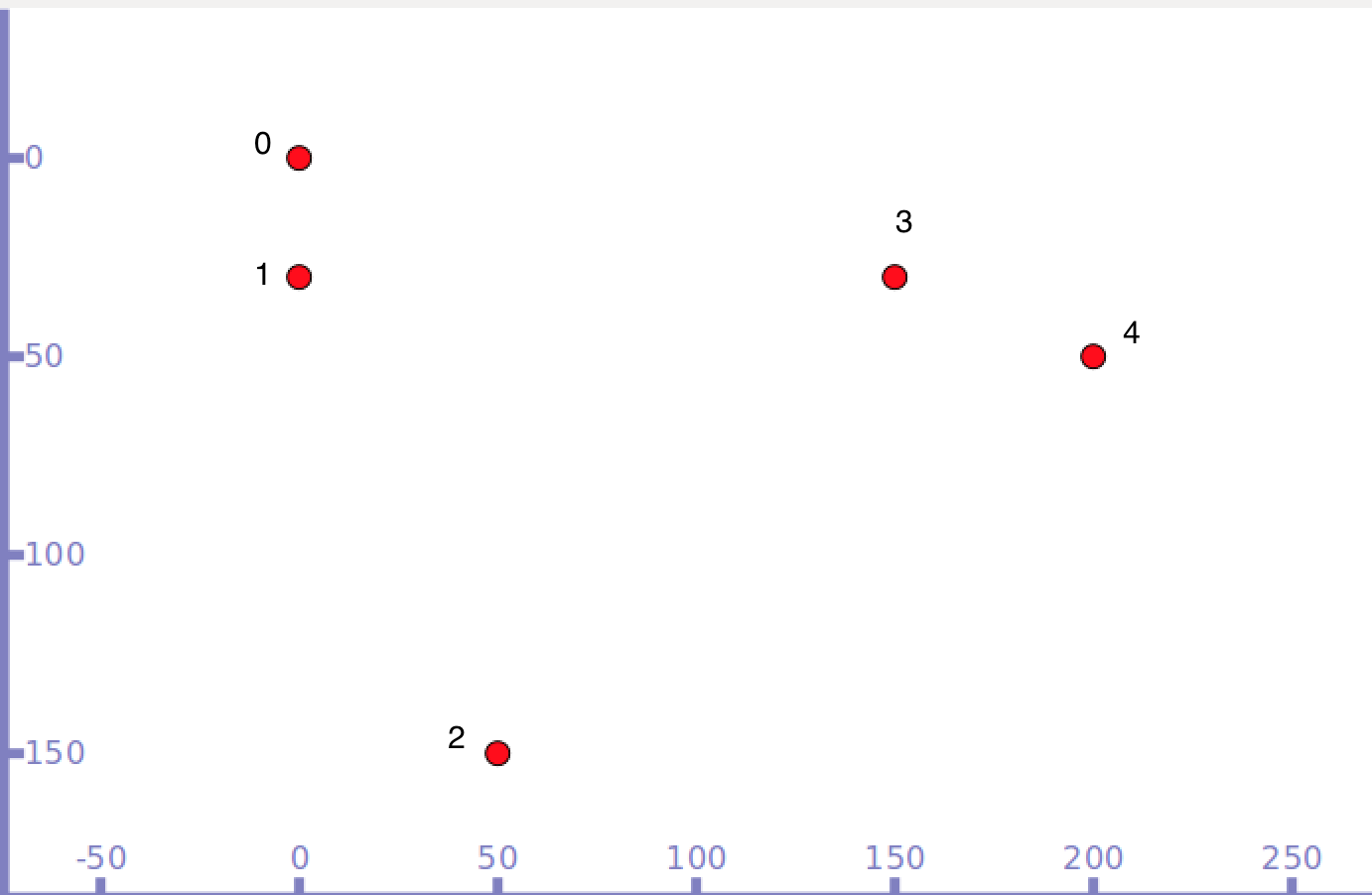}
  \caption{\small The 5 node simulation setup.}
  \label{Figure:5nodes}
\end{figure}

For the sake of simulation simplicity, we only work with the data packets of the torrent file. In the 5 nodes scenario (Figure~\ref{Figure:5nodes}), node 0 and node 4 start off with the complete set of data packets for the first and second torrent file (movie1, movie2), respectively. Node 2 wishes to download the first torrent file and node 1 wishes to download the second torrent file. Node 3 represents a pure forwarding node.

\begin{figure}[h]
  \centering
  \includegraphics[width=\columnwidth]{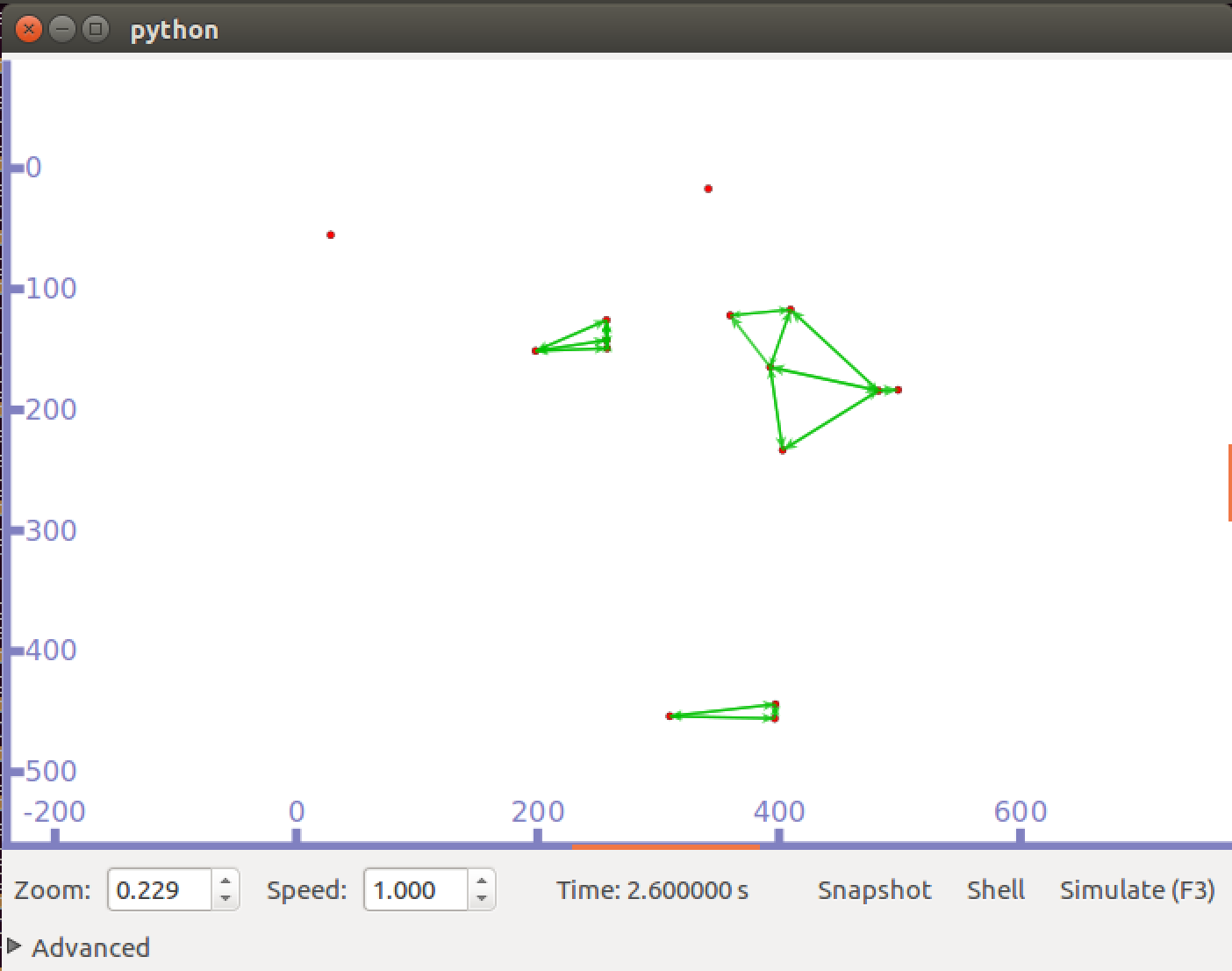}
  \caption{\small The main simulation setup.}
  \label{Figure:sim-setup}
\end{figure}

In Figure~\ref{Figure:sim-setup}, it illustrates another simulation used to test the implementation. Peers are randomly placed in the environment and move in a random direction every 20 seconds at a speed between 2 m/s and 10 m/s. Peers are bounded by the grid size to avoid peers moving out of the boundaries. Of the nodes placed in the simulation, one third of the peers wants the first torrent file and one third wants the second. There is one peer that has the complete set of data packets for each torrent file. The rest of the peers are pure forwarding nodes. For both simulations, different seeds can be used to ensure randomness for each trial.

\section{Conclusion \& Future Work}

This paper describes the design and implementation of multi-hop communication for nTorrent in a wireless ad hoc environment. With this design, we show that peers can still download their desired torrent files even while communicating with and through some nodes that do not run nTorrent and some that want other files.

In the future, we would want to run more simulations to mimic realistic situations and have more than just two torrent files. We would also like to increase the size of the torrent files to match real world files to see how well this design would work. Also, in the current ad hoc nTorrent implementation, we only work with the data packets for each torrent file. We can work on including the .torrent file and the file manifests into the ad hoc nTorrent application in the future.



\bibliographystyle{plain}
\bibliography{reference}

\end{document}